\documentclass[twocolumn,aps,showpacs,prb,epsf,graphics,psfig]{revtex4}
\usepackage{graphicx}
\usepackage{graphicx}
\usepackage{bbold}
\usepackage{color}
\usepackage[normalem]{ulem}
\usepackage{amssymb}
\usepackage{tikz}
\usepackage[version=3]{mhchem} 
\usepackage{amsmath}

\usepackage{lineno}

\begin{document}
\title{Differential tissue sparing of FLASH ultra high dose rates: an {\it in-silico} study}

\author{Ramin Abolfath$^{1,2,\dagger}$, Alexander Baikalov$^{3,4}$, Sohrab Rahvar$^2$, Alberto Fraile$^5$, Stefan Bartzsch$^{3,6}$, Emil Sch\"uler$^{1}$, Radhe Mohan$^{1}$}
\affiliation{
$^1$ Department of Radiation Physics and Oncology, University of Texas MD Anderson Cancer Center, Houston, TX, 75031, USA \\
$^2$ Physics Department, Sharif University of Technology, P.O.Box 11365-9161, Azadi Avenue, Tehran, Iran \\
$^3$ Technical University of Munich, Department of Physics, Garching, Germany \\
$^4$ Helmholtz Zentrum M\"unchen GmbH, German Research Center for Environmental Health, Institute of Radiation Medicine, Neuherberg, Germany \\
$^5$ Nuclear Futures Institute, Bangor University, Dean Street, Bangor, LL57 2DG, United Kingdom \\
$^6$ Technical University of Munich, School of Medicine and Klinikum rechts der Isar, Department of Radiation Oncology, Munich, Germany
}



\date{\today}
\begin{abstract}
{\bf Purpose}:
To propose a theory for the differential tissue sparing of FLASH ultra high dose rate (UHDR) through inter-track reaction-diffusion mechanism.

{\bf Methods}:
We calculate time-evolution of particle track-structures using a system of coupled reaction-diffusion equations on a random network designed for the molecular transport in porous and disorder media.
The network is representative of the intra- and inter-cellular diffusion channels in tissues. Spatial cellular heterogeneities over the scale of track spacing have been constructed by incorporating random fluctuations in the connectivity among the network sites.

{\bf Results}:
We demonstrate the occurrence of phase separation among the tracks as the complexity in intra- and inter-cellular structural increases.
The tracks evolve individually like isolated islands with negligible inter-track overlap at the strong limit of disorder as they propagate like localized waves in space, resembling the localized distribution of nano-plasmonic excitations in tumors, an analog of Anderson localization in quantum mechanics.
In contrast, in a homogeneous media and at the limit of weak disorder such as in water and normal tissues, the neighboring tracks melt into each other and form a percolated network of non-reactive species.
Thus, the extent of chemically active domains and their time evolution depends on tissue types such that
the spatio-temporal correlation among the chemical domains vanishes as the inter-cellular complexity of the tissues increases from normal to fractal-type malignancy.
Our model hypothesis on the induction of chemical species into isolated islands by the passage of particles in tumors and is consistent with the existence of isolated pockets of molecular oxygen in hypoxic conditions.

{\bf Conclusions}:
FLASH normal tissue sparing is a result of the interplay of the proximity of the tracks over intra- and inter-cellular landscape, a transition in the spatial distribution of chemical reactivity, and {\it molecular-crowding}.
In this context, insensitivities in the radiobiological responses of the tumors to the high dose rate at FLASH-UHDR are interpreted via a lack of geometrical correlation among isolated tracks.
The structural and geometrical complexities of cancerous cells prevent clustering of the tracks over a timescale that inter-track chemical reactivities presumably prevail in normal tissues.
This theoretical study presents proof of principle in a hypothetical scenario in which cellular complexity
influences dramatically the geometrical correlations of the track-structures. We provide a logical interpretation of
the experimentally observed differential FLASH-UHDR sparing of tissues. A series of systematic experiments on
radiolysis diffusibility and reactivity on actual normal and cancerous tissues must be carried out to classify the tissues potentially spared by FLASH-UHDR and verify our theory.

\end{abstract}

\maketitle
\section{Introduction}
The unique normal tissue sparing of FLASH ultra high dose rates (UHDR), i.e., 40 Gy/s and higher, has recently attracted considerable attention
[\onlinecite{Favaudon2014:STM,Montay-Gruel2018:RO,Vozenin2018:CCR,Montay-Gruel2019:PNAS,Buonanno2019:RO,Vozenin2019:RO,Arash2020:MP,Spitz2019:RO,Koch2019:RO,Abolfath2020:MP,Lai2021:PMB,Seco2021:MP,Abolfath2022:PMB,Baikalov2022,
Kusumoto2022:RR,Heuvel2022:PMB}].
The interpretation of the experimental data and the underlying microscopic mechanism are, however, under investigations and debates among researchers in the field of radiation therapy.

Among all theories proposed for interpretation of the experimental data (e.g., see Refs.
[\onlinecite{Favaudon2014:STM,Montay-Gruel2018:RO,Vozenin2018:CCR,Montay-Gruel2019:PNAS,Buonanno2019:RO,Vozenin2019:RO,Spitz2019:RO,Koch2019:RO,Lai2021:PMB}]), the authors of the present work have hypothesized transition between intra- and inter-track reactions as the major physical mechanism for differential biological responses of conventional dose rate (CDR) vs. FLASH-UHDR [\onlinecite{Abolfath2020:MP,Abolfath2022:PMB,Baikalov2022}].
In these models, the time evolution of radiolysis products are assumed to propagate in homogeneous and unform medium, regardless of tissue types.
Thus the presented models lack differentiation with respect to tissue types upon exposure at FLASH-UHDR.

A series of systematic experiments recently conducted and published by Kusumoto {\em et al.} [\onlinecite{Kusumoto2022:RR}]
on measurement of chemical yields of 7-Hydroxy-Coumarin-3-Carboxylic acid in solutions irradiated by proton- and carbon-ion beams at UHDRs.
These experimental studies have revealed evidences in favor of inter-track coupling hypothesis, originally predicted
by performing molecular dynamics simulations of track-track chemical interaction.
The results of the simulations, presented in Ref. [\onlinecite{Abolfath2020:MP}], have led to interpretation of {\it molecular-crowding} in population of reactive oxygen species (ROS) and formation of agglomerates in form of non-reactive oxygen species (NROS), consistent with recent observations reported in Ref. [\onlinecite{Kusumoto2022:RR}].

In this work, we extend our model-calculation on the same physical grounds as in Refs. [\onlinecite{Abolfath2020:MP,Abolfath2022:PMB,Baikalov2022}] and take into account the cellular structure of normal and cancerous tissues at a coarse-grained scale and incorporate the tissues differential sparing of FLASH-UHDR to extend predictions and outcome of our inter-track coupling hypothesis on tissue types, consistent with empirical observations.
We propose theoretically an interplay between the rates associated with diffusion and recombination of ions,
and predict occurrence of intra- to inter-track transitions as a function of cellular structure and tissue types,
i.e., from tumors to normal tissues respectively.

\subsection{Terminology}
In a nutshell, passage of a high energy particle (electron, proton, or heavier charged particles) in matter leaves a linear dynamical foot-print from cylindrically symmetric (isotropic) exchange of energy with electrons and nuclei constituting molecular structures.
This linear structure and its branches is known as a particle {\it track}.

A single track is a random collection of sharply spatio-temporal distribution of non-ionized and ionized excitations, with a varying nano-scale diameter which depends on the particle kinetic energy that determines the magnitude of energy exchange.
Due to quantum electrodynamic (QED) nature of energy exchange, the excitations are created within atto-seconds time delay after passage of particles.

Immediately after their creations, molecular excitations and ions undergo decay processes.
The relaxation time associated with the decay of excitations (including recombination of mobile ions into various types of stable  products and chemical species) are much longer than their generation time.
Because the excited molecules and ions are mobile in cellular structures, they decay at the same time as they diffuse away from the center of track.

Presence of high concentration of localized excitonic energy of molecules, surrounding mobilized ions, induces an explosive irreversible flow of thermal energy to ion-species which in turn, theoretically, enhances significantly the effective diffusion constant of ions.
Ions move randomly along the radial direction away from the hot core of the tracks with a thermally boosted kinetic energy that generates shock-waves [\onlinecite{Abolfath2022:PMB,Friis2021:PRE,Friis2020:JCC,Fraile2019:JCP}].
They asymptotically lose their kinetic energy and fall into cold diffusion at thermally equilibrium condition because of collisions and exchange of energy with the molecules in the environment.
Eventually ions rest at room-temperature with transformed chemical composition.

The core temperature of a track depends on the particle type and its linear energy loss per length (LET).
It can go up to several thousands of kelvin for heavy charged particles [\onlinecite{Abolfath2022:PMB}].
Throughout this process, biological damages to the host cellular structure take place as ions interact chemically with bio-molecules, such as DNA.

A typical radiotherapy beam of particles form a random distribution of expanding and decaying tracks in targeted (tumors) and untargeted (normal-tissues) volumes.
As pointed out previously, similar to a single track configuration, the tracks induced by a beam of particles initially expand individually via a time-dependent diffusion mechanism as they decay because of deexcitation and ion-recombination processes.
The time evolution of such ensemble of tracks can be reduced to a single track if the geometrical overlap among the tracks is negligible.
We refer to this limit as ``independent track structure".
Conversely, ``strongly correlated track structure" can be anticipated at a limit where
the process of inter-track ion exchanges, chemical transformation and recombination takes place simultaneously due to destructive interference of sufficiently close tracks, a {\it molecular-crowding} phenomenon.
We therefore refer to these two distinguishable classes of chemical exchange mechanisms as intra- and inter-track states.

The transition between intra- and inter-track recombination depends on the dose and dose rate.
More precisely, the higher beam intensity (the number of particles entering a unit area per unit time),
the higher the compactness of particles in a time-interval hitting the target.
It allows the tracks to be closer to each other within an interval of time.
In this limits, the overlap probability among the tracks prior to their annihilation becomes significant.
Under certain conditions a transition from the intra-track to inter-track reaction has been predicted.
In our recent publications [\onlinecite{Abolfath2020:MP,Abolfath2022:PMB,Baikalov2022}], the latter has been hypothesized as physical mechanism for FLASH ultra-high-dose-rate (UHDR).

\section{Materials and Methods}
\label{Sec_MM}
\subsection{Track spacing}
Passage of high energy particles in cells, tissues, or water-equivalent materials generate highly localized tracks within nanoscopic scale in a very short period of time.
At UHDR, the instantaneous track cross sectional (two dimensional) distribution depends on the total dose delivered to the tissue volume, thus it is a function of particle fluence, in addition to the particle type, energy, LET and depth.
We refer the interested readers to our recent publication [\onlinecite{Baikalov2022}] on the details of the track calculation and the mean lateral spacing.

\subsection{Reaction-diffusion model}
Right after calculation of the deposition of dose at UHDR, with a packed lateral distribution and given three-dimensional landscape of the tracks, we carry out a second calculation based on a system of coupled reaction-diffusion equations to simulate transport of chemical products generated by ionizing radiation in a cellular medium.
We focus on the calculation of the ratio of intra- and inter-track chemical interactions and geometrical correlations, e.g., their overlaps.

In this model, the radiation induced chemicals are concentrated in a core of cylindrically symmetric body / cloud of track structures.
The mathematical details of our model calculation with analytical solutions for time-evolution of a single-track are given in the Appendix \ref{AppNumApproach}.

As a representative of reactive oxygen species that causes DNA damage, we consider OH-radicals.
OH-radicals are known to diffuse through cellular space and react with biomolecules including DNA.

If cells were uniform and homogeneous, like in liquid water, the diffusion of ions induced by radiation took place like in an ordered medium.
The current  models in radiobiology, however, do not take into account intra- and inter-cellular inhomogeneities in diffusion of radiolysis products.

As a first step in proof-of-principle and to demonstrate the effects of cellular structures and textures on interpretation of the
tissue-sparing of FLASH-UHDR,
we consider two types of mediums to study transport of chemical species in typical normal and tumor cells/tissues.
Because of substantial differences in intra- and inter-cellular structure and chemical compositions of tumor vs. normal cells,
we solve reaction-diffusion equations in a homogenous and isotropic medium, similar to liquid water, as a representative of normal cells/tissues and in a heterogenous fractal-type porous and disordered medium for tumors
[\onlinecite{Klein2013:NL,Thiagarajah2006:NM}].

It is necessary to comment on the details of cellular structures such as exact locations of various organelles.
In our model, the detailed information on cellular mass inhomogeneities are averaged out with respect to the track locations.
Because in a typical radiotherapy beam of particles, track locations are randomly distributed among another random distribution of the cells in tumors and normal-tissues, a compound distribution as has been used in formulation of theory of dual radiation action (TDRA)~[\onlinecite{Kellerer2012:RR,Abolfath2019:EPJD}].


\section{Results}
\begin{figure}
\begin{center}
\includegraphics[width=1.0\linewidth]{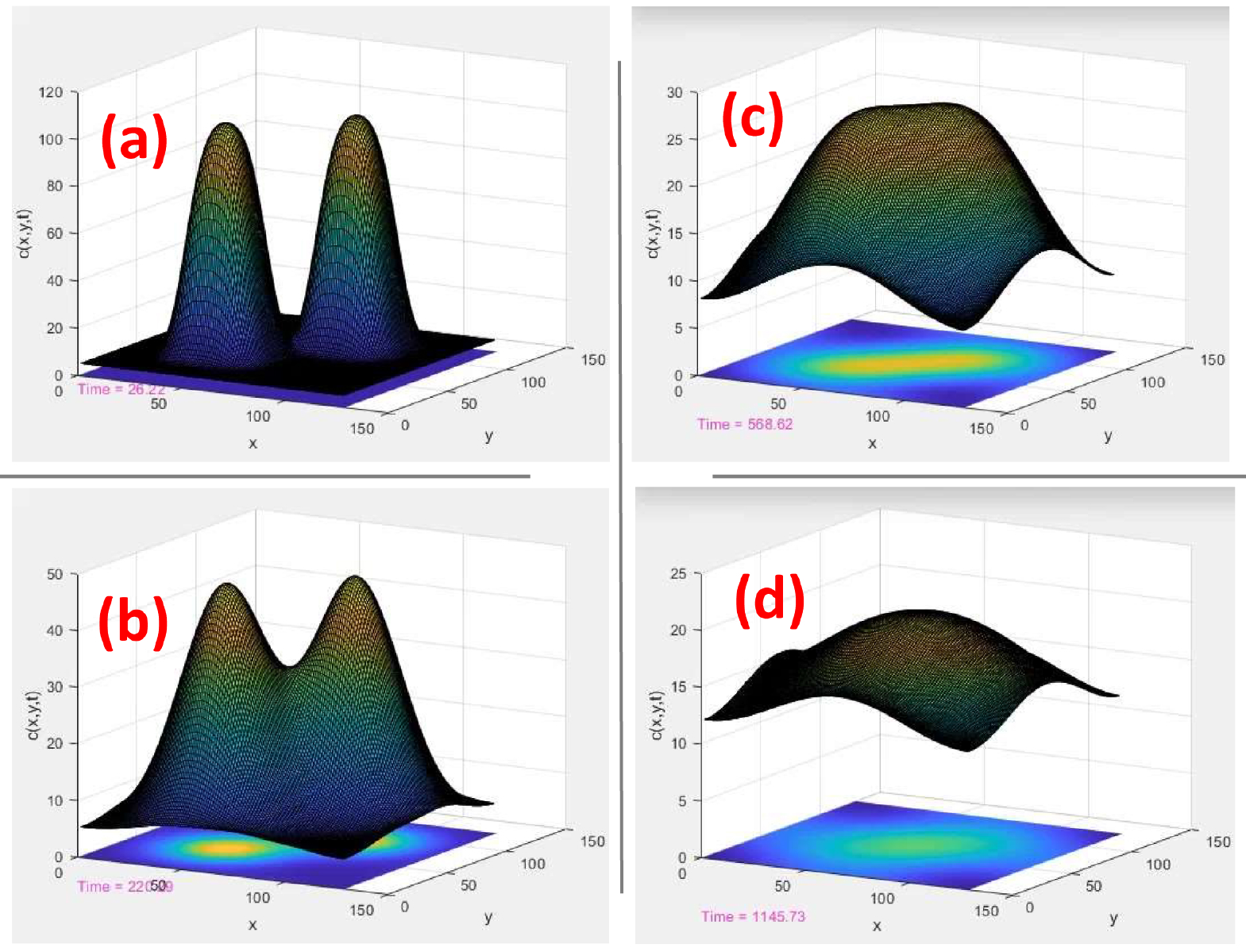}\\ 
\noindent
\caption{
Time evolution of two tracks, $u_i(x,y)$ and $u_j(x,y)$, in a homogeneous and uniform medium.
$c(x,y) = u_i(x,y) + u_j(x,y)$ is total density of ROS calculated by superposition of individual ROS's.
$(x,y)$ are the planner coordinates of the plane perpendicular to the axis of cylindrical tracks.
}
\label{fig1}
\end{center}\vspace{-0.5cm}
\end{figure}

\begin{figure}
\begin{center}
\includegraphics[width=1.0\linewidth]{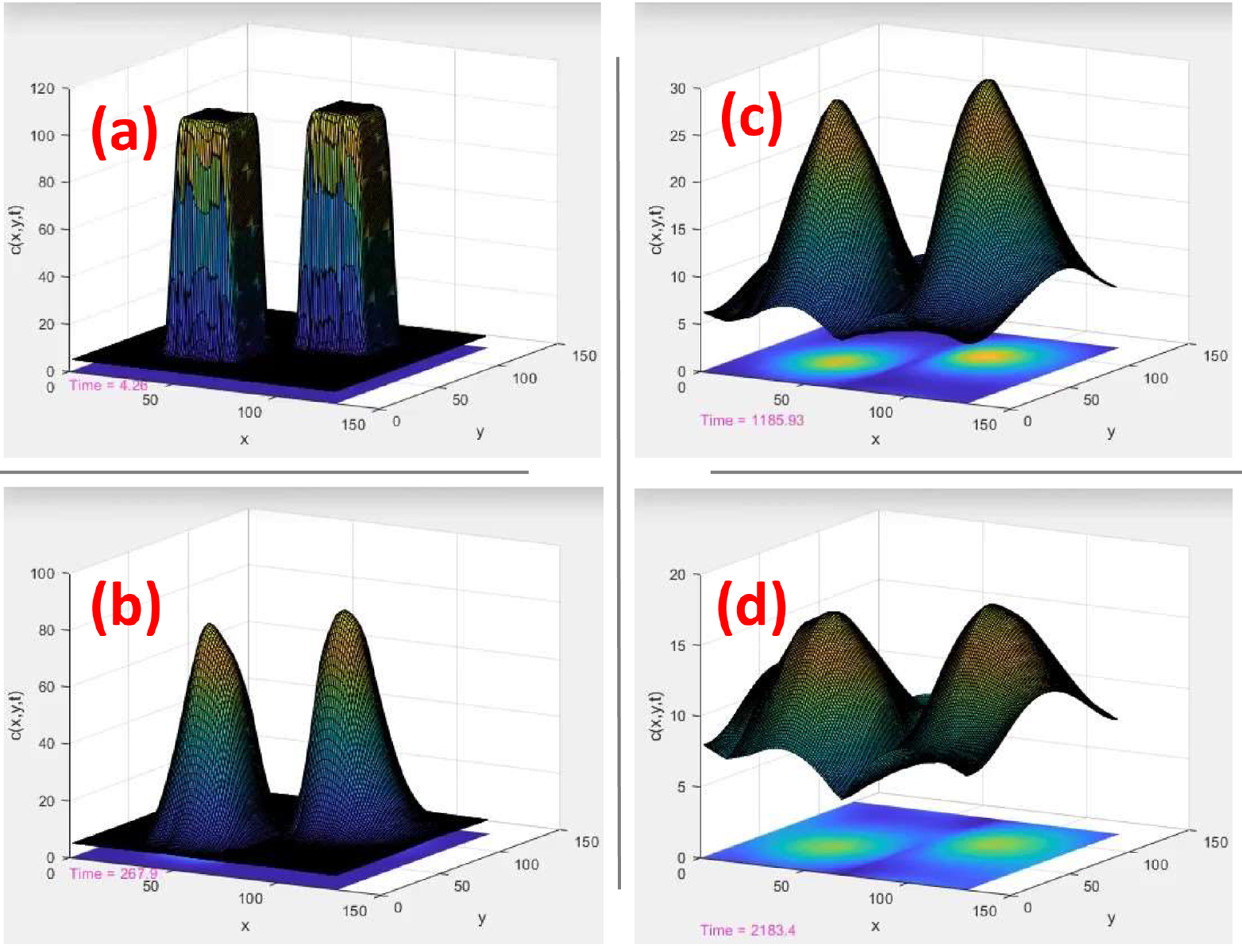}\\ 
\noindent
\caption{
Time evolution of two tracks in porous medium with random connectivity among the diffusion sites.
Similar to Fig. \ref{fig1}, $c(x,y) = u_i(x,y) + u_j(x,y)$.
}
\label{fig2}
\end{center}\vspace{-0.5cm}
\end{figure}

Figs. (\ref{fig1}) and (\ref{fig2}) present the time evolution of two tracks simultaneously started in two cylindrically symmetric clouds of ionization with radius $w$.
The real-time motion of these tracks are available online.

In Fig. (\ref{fig1}-a, -b, -c, -d), a solution of 2D reaction-diffusion equation as a function of time was calculated in a homogenous and uniform medium such as in water.
As shown, two cylindrical tracks evolve initially into two uncorrelated Gaussian probability distribution functions (PDFs) with centers located at $\vec{r}_i$ and $\vec{r}_j$ before they collapse together, where
\begin{eqnarray}
u_i(\vec{r}, t) = \frac{e^{-\frac{|\vec{r}-\vec{r}_i|^2}{4D_f(t - t_i)} - k_1(t-t_i)}}{4\pi D_f (t - t_i)},
\label{eq1}
\end{eqnarray}
and
\begin{eqnarray}
u_j(\vec{r}, t) = \frac{e^{-\frac{|\vec{r}-\vec{r}_j|^2}{4D_f(t - t_j)} - k_1(t-t_j)}}{4\pi D_f (t - t_j)}.
\label{eq2}
\end{eqnarray}
Here $D_f$ and $k_1$ are the diffusion constant and reaction rates, respectively.
Our approach on numerical calculation of the time-dependent solutions of the diffusion equation subjected to a cylindrically symmetric initial condition and fitting to Gaussian functions at distances away from the cylinder can be found in Appendix \ref{AppV}.

Without loss of generality, to illustrate the effects of tissue types, we considered the creation time of tracks $t_i = t_j$ in these simulations.
This is a condition that approximately fulfill the time sequence of the track inductions at UHDR.
Note that in general, the temporal distribution of the tracks, hence their relative time elapse, depends on the dose rate.
However at UHDR, we can neglect the time elapse among the tracks in comparison with other time scales involved in the present reaction-diffusion model.

As the simulation time proceeds in Fig. (\ref{fig1}), from (a) to (d), two Gaussians merge together and form an elongated single PDF.
The geometrical overlap of two Gaussians can be calculated analytically
\begin{eqnarray}
\langle u_i | u_j \rangle(t) &=& \int d\vec{r} u_i(\vec{r}, t) u_j(\vec{r}, t) \nonumber \\
&=&
\frac{e^{-\frac{|\vec{r}_i-\vec{r}_j|^2}{8D_f t} - 2k_1 t}}{8\pi D_f t}.
\label{eq3}
\end{eqnarray}
As two Gaussians combine together, like melting two droplets into a single droplet, the diffusion slows down in the overlap area.
Instead, the diffusion carries out with a rate calculated by Eq. (\ref{eq3}) from the periphery of combined-Gaussians to outside.

In Fig. (\ref{fig2}-a, -b, -c, -d), we have calculated a solution of a reaction-diffusion equation with identical initial condition as in Fig. (\ref{fig1}-a, -b, -c, -d) except the calculation has been performed on a network with random connectivity between the neighboring sites to mimic the geometrical disorder of tumor cells with strong inhomogeneity and/or fractal-type porosity.

A series of connectivity probabilities, $p$, have been drawn from a unform distribution within the interval of zero and one and subsequently have convoluted to diffusion constant, $D_f$, for each diffusion site in the network.
Although the reaction rate, $k_1$, can be considered another random variable, but we have kept it constant, the same value as in the simulation shown in Fig. (\ref{fig1}) to isolate the effects of diffusion.
Note that a special case of $p=1$ describes transport of ions on a homogeneous network with uniform connectivity that links nearest neighbor sites, corresponding to the kinetics of ions among normal cells with the results depicted in Fig. (\ref{fig1}).

At every simulation time step, the diffusing ions select randomly one of its nearest neighbor sites.
If the move to that site is allowed with probability, $p$, the ion moves one step outward.
Otherwise the ion stays on the initial site with probability $1 - p$.
The diffusion constant of such Brownian particle can be calculated by Einstein relation, $\langle r^2 \rangle = D_f t$.
Here $\vec{r}$ is Euclidean distance that measures how far the particle has moved randomly away from the center of coordinates where it was created.
Above the network percolation threshold ($p > p_c$), the Brownian motion can find at least one trajectory to cross the entire system, $D_f = \langle\langle r^2 \rangle\rangle / t$, otherwise $D_f = 0$ (including at the percolation point, $p = p_c$).
Note that $\langle\langle \vec{r} \rangle\rangle = 0$ because of unbiased random-walk considered in these simulations.
For a given $p$, $\langle\langle \cdots \rangle\rangle$ represents double averaging, i.e., random walk averaging subjected to a specific network configuration, followed by ensemble averaging over a large number random network configurations.
For a review on percolation theory and complex networks see, e.g., Ref. [\onlinecite{Li2021:PR}].

For a perfect network where $p=1$, $D_f$ is the maximum.
It decays continuously to lower diffusion values for $p_c \leq p \leq 1$ and vanishes at $p = p_c$.
$D_f$ remains zero within $p \leq p_c$.
Note that close to $p=p_c$ (from the above), the clusters in the network form a fractal-type structure with a Hausdorff dimension that is a measure of the tissue / cell roughness, or more specifically, their fractal dimension.
Below $p_c$, the clusters are isolated thus the diffusion through entire tissue / cell stops to occur.

The time and length scales in Figs. (\ref{fig1}) and (\ref{fig2}) have been chosen based on the conventional values of the diffusion constants.
To simulate expansion of a track of OH-radicals at thermal equilibrium with environment at room temperature and using an empirical value $D_f = 4.3 \times 10^{-9} m^2/s = 0.43 \AA^2/ps$ [\onlinecite{Abolfath2022:PMB}], we divide the square sides of the computational boxes into steps with 0.1 nm length.
In these calculations the time advances via 0.1 ps intervals to fulfill the Nyquist sampling theorem in signal processing in which the simulation time steps are required to be half or less of the period of the quickest dynamics.
Accordingly, such length-scales set the lateral sides of the computational boxes in Figs. (\ref{fig1}) and (\ref{fig2}) to 13 nm. 
The running time of these simulations have terminated at $0.5 \mu s$ with no significant differences from the times corresponding to Figs. (\ref{fig1}d) and (\ref{fig2}d).

\begin{figure}
\begin{center}
\includegraphics[width=1.0\linewidth]{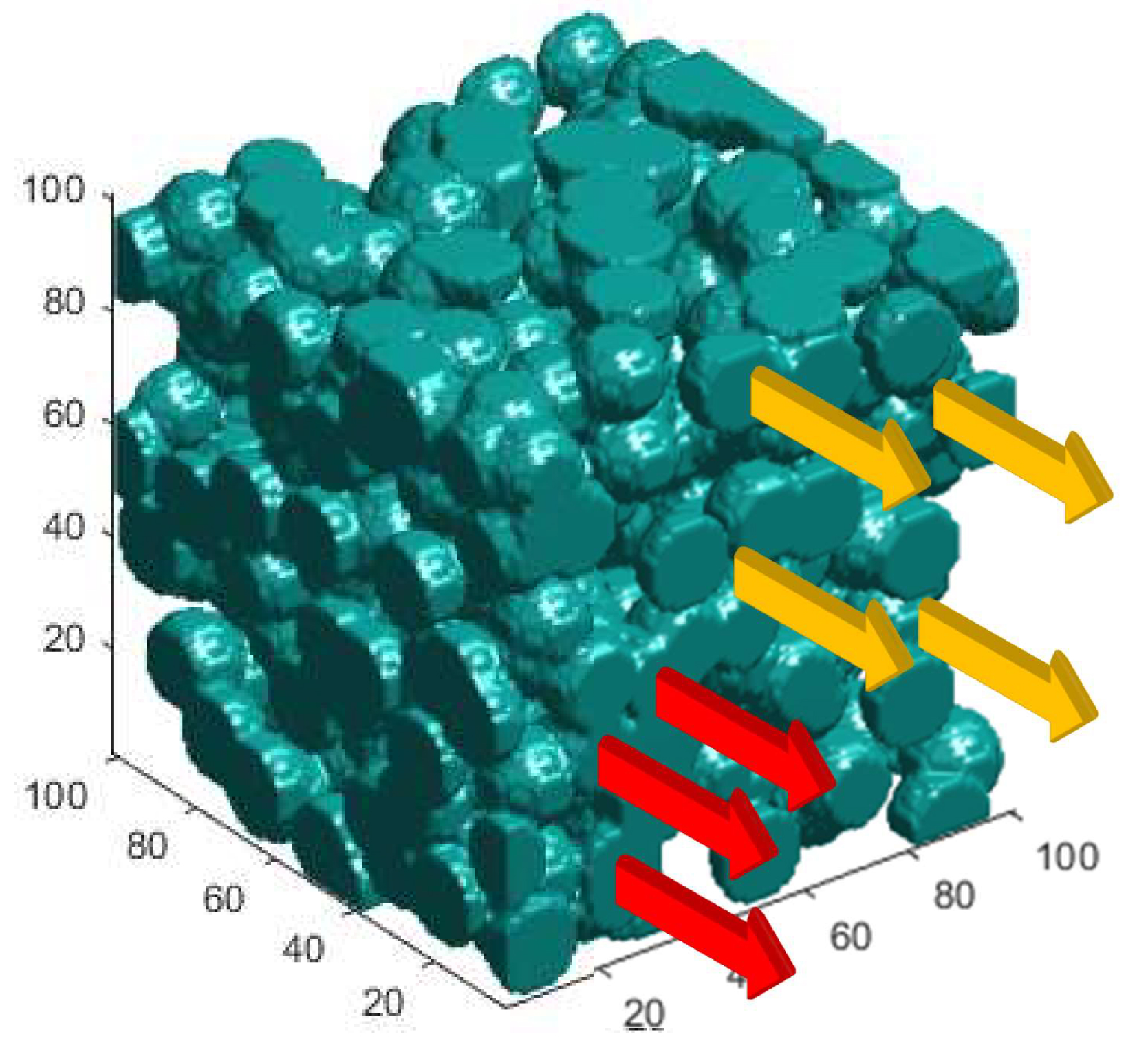}\\ 
\noindent
\caption{
Schematic sketch of tracks passing through a porous media, a representative of heterogeneities in tissues.
Shown in green are clusters with diffusion constant $D_f$.
The diffusion is forbidden in void space among the clusters.
Red arrows are representative of interacting (mediated by diffusion) tracks as they pass through a single cluster, hence they are connected tracks.
Orange arrows passing through isolated islands and are representative of non-interacting tracks.
}
\label{fig3}
\end{center}\vspace{-0.5cm}
\end{figure}

The overlap between two adjacent tracks is expected to happen at time scale $t = \ell^2 / D_f$ if the relevant length scale for diffusion, i.e., the diffusion length, $\langle\langle r^2 \rangle\rangle^{1/2}$, becomes comparable to inter-track spacings, $\ell$.
Even below the percolation limit, $p < p_c$, two tracks can be connected through intra-cluster diffusion channels if two or more tracks pass through a single cluster.
Another interesting construction of a system of tracks and isolated clusters can be represented by two neighboring tracks that pass through two separated and disconnected clusters with no diffusion channel between them.
This combination corresponds to a non-interacting track configuration as shown in Fig. \ref{fig3}.
In this figure, tracks with different color codes are designated based on their classifications as interacting (red) and non-interacting (orange).
The underlying porous media, representing a typical tumor tissue, is depicted in green where the diffusion can be carried out.
The clusters are separated by clear voids, the space where the diffusion is forbidden.

Collection of configurations of a system of tracks and tissue-clusters under the condition, $p < p_c$, some with finite $D_f$, combined with vanishing $D_f$, lead to system of tracks with lower effective interaction compared with tissues under the condition, $p > p_c$, where all clusters are connected.
The former represents tumors and the latter represents normal tissues.
The problem as such is interesting from a mathematical point of view as it describes time evolution of percolating tracks mediated through diffusion channels subjected to percolation of the underlying medium, cellular structures and tissues, i.e., a compound percolation system.

Based on the discussion above, note that the time evolution of the diffusion process shown in Fig. (\ref{fig2}-a, -b, -c, -d), is one of the configurations of the network corresponding to $p$ close to $p_c$.
Similar configurations with the small mean value in diffusion constant, $D_f$, can be generated by repeating the same calculation, as in Fig. (\ref{fig2}) but starting with different random seeds.

At FLASH-UHDR conditions, if the correlation length in the network connectivity, $\xi$, that is a measure of cluster size, is smaller than the mean inter-track distances, the diffusion effectively do not occur to the extent of track spacings thus the response of tissue falls into the class of isolated / single track states.
This is a scenario the percolation theory predicts for typical tumor cells / tissues irradiated by a source of FLASH-UHDR.

As can be seen clearly from these two simulations, the effect of randomness in connectivity among the diffusion channels is to localize the tracks where the cell/tissue responses is insensitive to the time-elapse among the tracks, simply because of negligible inter-track overlaps. Hence the tissues with strong porosities and disordered in their diffusion channels (either normal or cancerous), under radiation must exhibit insensitivity to the dose rate, the same phenomenon observed empirically from the tumors under FLASH-UHDR.

\section{Discussion}
Owing to their highly chemical reactivities, particle tracks, a nanoscopic cloud of mobile plasmonics, are known to be major sources of indirect DNA damage, following by chromosome sub-lethal injury, and cell death.
Their biological pathways are triggered by physical and chemical processes that have been well formulated by TDRA~[\onlinecite{Kellerer2012:RR,Abolfath2019:EPJD}].
This process occurs during diffusion of chemical species, constituent of the track structure.
It is, therefore, crucial to capture essential underlying mechanisms of reaction-diffusion processes of chemical species in cells to properly model the differential aspects of tissue sparing at FLASH-UHDR.

In that regards, it is important to remind that these mobile chemical species are initially embedded inside of a larger shell of a localized and hot cloud of non-ionizing molecular excitations.
Monte Carlo (MC) simulations of track structures, e.g., Geant4-DNA [\onlinecite{Incerti2010:IJMSSC}], have shown that up to 45\% of a particle kinetic energy transfers to generate thermal spikes and the rest to generate  mobile ion-species.
The thermal spikes in form of hot molecular excitations may also contribute to the ionization processes of the medium before they dissipate to thermo-acoustic waves.

Because of exchange coupling of ion-species with non-ion excitations through thermal exchange interaction, mediated by acoustic phonons, the ion-species initially gain large kinetic energy and their random diffusive motion is thermally boosted.
Such coupled system initially propagates like blasts of explosive at nano-scales.
Because they move in a substantially strong dissipative cellular media, they quickly lose their kinetic energy and asymptotically decay into lower state of diffusion at the thermal equilibrium matched with the cellular (room) temperature [\onlinecite{Abolfath2022:PMB}].

In recent years several techniques have been developed to measure molecular diffusion in cellular environments [\onlinecite{Klein2013:NL,Thiagarajah2006:NM,Dix2008:ARB,Mourao2014:BJ,Cross2007:NN,Zink2004:NRC}].
In general the diffusion constant substantially depends on the size of molecules, the roughness of the inter- and intra-cellular structures and chemical compositions and texture of the cells.
The cellular texture varies in a range of uniform and homogeneous to strongly disordered, exhibiting fractal geometries, as in tumor cells [\onlinecite{Klein2013:NL,Thiagarajah2006:NM}].
The latter mechanism bounds the range of molecular random-walks and blocks Brownian motion of chemical pathways below the percolation limit of the diffusion at $p = p_c$ associated with the underlying fractal dimensionality and self-similarity of the cellular structure.
Spite of these reports and observations, there is no study in the radio-biological literature to address the effects of cell types on diffusion of ion-species.
In particular that all models neglect the differences in heterogeneities among tissues and consider all cell types equivalent of uniform and homogenous liquid water.
The aim of this study is to highlight qualitatively the tissue heterogeneities in modeling inter-track coupling at FLASH-UHDR.
More accurate models require incorporation of cellular internal and external structures in calculation of $D_f$, beyond the scope of this paper.



The underlying physical processes of tissue response to radiation dose rate, including differential biological responses of various tissues, either normal or malignant, can be formulated throughout overlap among tracks.
In this model the tissue biological responses are categorized based on the geometrical correlation and collective evolution of the tracks.
In a single fraction, tracks with negligible overlaps do not lead to a physio-chemical responses sensitive to radiation dose-rate.
Thus the typical tumor responses fit to a class of uncorrelated and evolutionary single tracks, the dominant intra-track effects.
In contrary, normal tissue responses can fall into another class of collective {\it chemical-crowding} of the correlated tracks where inter-track effects are dominant.
The transition between inter- and intra-track reaction-diffusion processes are responsible respectively for these two seemingly distinguished behaviors among tissues.

The solutions of the coupled partial differential equations of two separate tracks were initially created at two separate positions are depicted in Figs. (\ref{fig1}) and (\ref{fig2}).
An underlying network among the reaction-diffusion sites have been considered to model the diffusion channels in tissues.
In this model, a tissue is a network with random connectivity among the sites.
In Fig. (\ref{fig1}), a network with uniform and homogeneous connectivity has been considered to represent normal tissues.
In Fig. (\ref{fig2}), a random network defined by a random connectivity is a representative of cancerous tissues identified to behave like fractals at the percolation threshold, $p = p_c$, the point where the diffusion channels are blocked due to emergence of the isolated islands.

The results shown in Fig. (\ref{fig1}) illustrate the role of tissue texture in forming overlaps among tracks as a function of time.
In Fig. (\ref{fig2}), randomness in diffusion channels, that is unique to transport through porous and disordered structures, limits the range of diffusion, thus the tracks evolve individually like isolated islands with negligible overlap.
This is consistent with the scaling theory of percolation and localization of thermal waves / Schrodinger equation (i.e., Anderson localization).

\begin{figure}
\begin{center}
\includegraphics[width=1.0\linewidth]{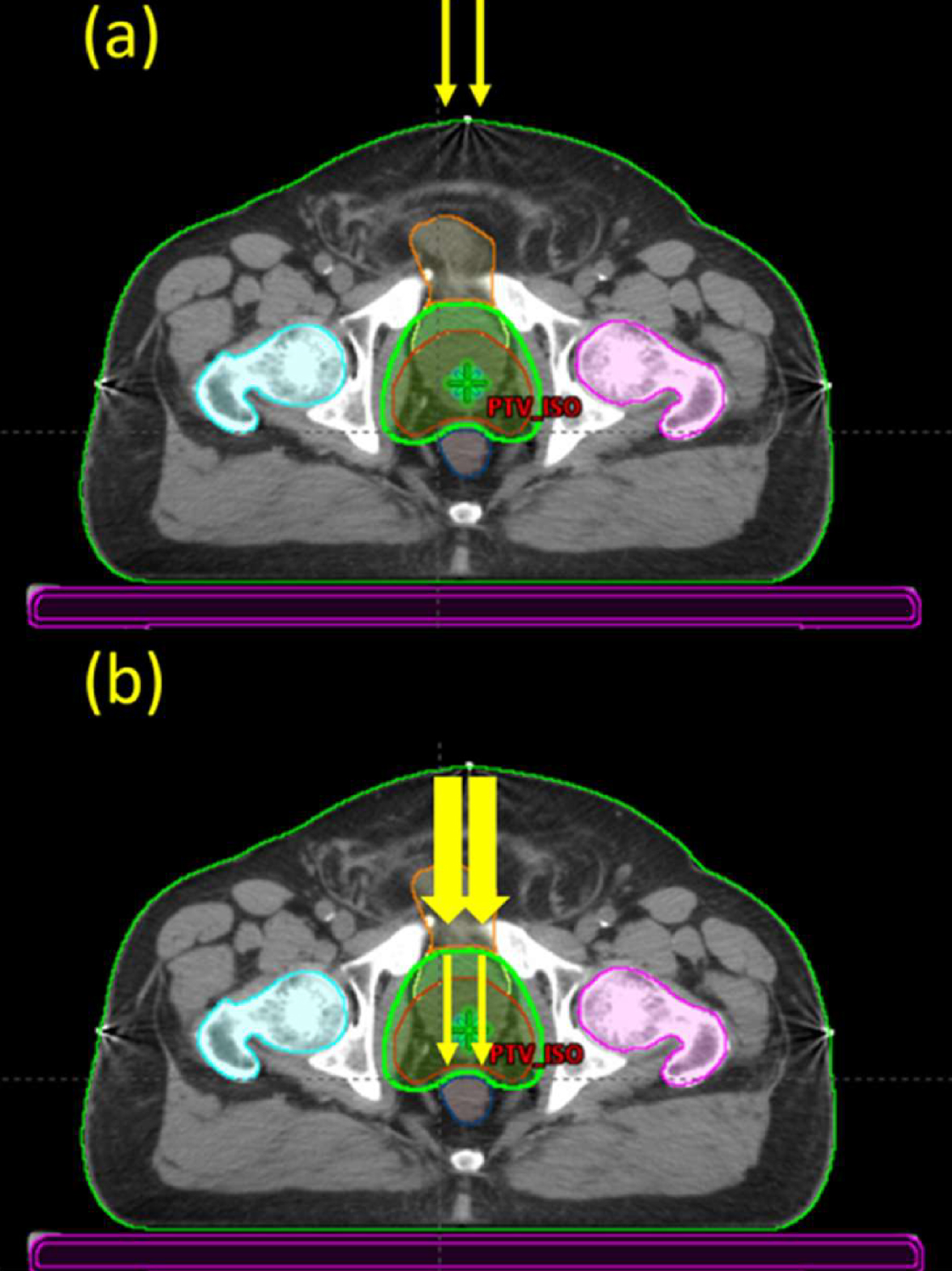}\\ 
\noindent
\caption{
Schematic sketch of diffusion expansion of two particle tracks in air prior to (a) and after (b) entering the patient body.
The width of arrows in normal tissues (thicker arrows) and in tumor (thinner arrows in prostate) tends to sketch the extents of inter-track overlaps in different tissue types.
In normal tissues, the expanded tracks allow inter-track overlaps while in tumor tissues the localized tracks yield negligible inter-track overlaps.
At given dose where the number of tracks (shown by arrows) is given, lack of inter-track overlaps justifies insensitivities of radiobiological responses to dose rates.
}
\label{fig4}
\end{center}\vspace{-0.5cm}
\end{figure}

Fig.(\ref{fig4}) (a) and (b) show schematically sketch of two beamlets prior and after entering the patient's body respectively.
The diffusive expansion of the beamlet tracks in normal tissues, depicted by the thicker arrows, and in tumor, depicted by thinner arrows in prostate, are seen.
At a given time after entering the beamlets, they expand more rapidly in normal tissue because of higher diffusibility compared to two isolated beamlets in tumor.
The larger expansion of tracks in normal yield higher overlaps.


Note that lowering the diffusion constant without incorporating the randomness in the network connectivity does not lead to localization of Gaussian PDFs as the absolute value of diffusion constant does not change the overall effect in inter-track evolution and their overlap, i.e., to block emerging two tracks together.
More precisely, the time evolution of the diffusion equation is invariant under the scaling of the diffusion constant.
A simultaneous scaling of diffusion length and time shows a similar trend in the tracks geometrical overlaps.
However, with constant intra- and inter-tracks reaction rates this scaling rule breaks down, unless we scale them simultaneously.

Finally, for the interested readers we remark that the track-localization observed in these simulations that is consistent with the percolation theory of diffusion on porous and disorder media has been extensively studies in the context of semiconductor physics.
The phenomenon known as the Anderson localization [\onlinecite{Lee1985:RMP}], has
extensively studied quantum mechanically to describe the metal-insulator transitions in condensed matter and solid state physics.
Here we map the normal and tumor tissues problem to similar transition between metals (where conduction electrons are in the extended states) and insulators (where conduction electron form pockets of localized states).
We suggest the mechanism modeled in these computer simulations interpret the empirically observed tissue-sparing of FLASH-UHDR for the first time.
This as a hypothesis alongside with the differential antioxidant concentrations or differential oxygen concentrations, currently under investigation.

\section{Conclusion}
This theoretical study aims to present a model calculation based on reaction-diffusion of reactive-species induced by ionizing radiation and point to possible transitions in the molecular-crowding of the track structures.
In this work we have presented a qualitative but algorithmic scenario to classify the clinical end points associated with the dose rate, such as differential biological responses of various tissues, either, normal or malignant, in a unified formulation throughout the overlap among tracks.
Tracks with negligible overlaps do not lead to a physio-chemical response sensitive to radiation dose-rate.
The underlying biological response of such effects stem from geometrical correlation and collective evolution of the tracks.
For the first time, our hypothesis explains the differential sparing effects observed empirically at FLASH-UHDR.
A systematic experimental cell/tissue database must be generated to validate the hypothesis presented in this work.
\section{Appendix: Numerical approach}
\label{AppNumApproach}
The rate equations proposed in this work, describe reactive oxygen species (ROS) aggregation and formation of non-reactive oxygen species (NROS) agglomerates such as transformation of a pair of OH to stable compounds such as H$_2$O$_2$ or transient and metastable complexes of OH$\cdots$OH.
We introduce two dynamical variables $u(\vec{r}, t)$ and $v(\vec{r}, t)$ and propose a system of coupled reaction-diffusion equations, denoting $u=$[OH] and $v=[$H$_2$O$_2]$.
These variables represent fast (ROS) and slow (NROS) moving species.

Conversion of ROS (OH) to NROS (H$_2$O$_2$) and vice versa can be described by the following rate equations
\begin{eqnarray}
\frac{\partial u}{\partial t} &=& G + \nabla \cdot\left(D_f(\vec{r}) \nabla u\right) - k_1 u + k_2 v - 2k_3 u^2 - k_{12} u v, \nonumber \\ &&
\label{Aeq1a}
\end{eqnarray}
\begin{eqnarray}
\frac{\partial v}{\partial t} = k_1 u - k_2 v + k_3 u^2.
\label{Aeq1b}
\end{eqnarray}
Here $G(\vec{r},t)$ and $D_f$ represent dose rate and diffusion constant of the fast moving species (neglecting the diffusion of slow moving species), and $k_1, k_2, k_3, k_{12}$ are reaction rate constants.

For a homogenous and uniform system, $D_f$ is a constant, hence in Eq. (\ref{Aeq1a}) we can substitute $D_f \nabla^2 u$ for $\nabla \cdot\left(D_f(\vec{r}) \nabla u\right)$.
In the following, we calculate analytical solutions of Eqs. (\ref{Aeq1a}) and (\ref{Aeq1b}), considering $D_f$ a constant.
However, for random networks considered in this work, $D_f$ is a function of position, $\vec{r}$.
In this case we calculate solutions of Eqs. (\ref{Aeq1a}) and (\ref{Aeq1b}) numerically.

Eqs. (\ref{Aeq1a}) and (\ref{Aeq1b}) are generalization of ROS-NROS rate equations introduced by Eqs.(1) and (2), in Ref.
[\onlinecite{Abolfath2020:MP}], where the “non-linearities” in the rate equations have shown the dominance of NROS at UHDRs.
Note that in the current work, we have added thermal diffusion and steady state decay terms ($k_1$ and $k_2$) where
in the absence of linear terms, $D_f = k_1 = k_2 =0$, we can recover Eqs. (1) and (2) in Ref. [\onlinecite{Abolfath2020:MP}] (after substituting the variables $N_1$ and $N_2$ for $u$ and $v$).


The numerical values of the rate constants are available in MC codes such as TOPAS n-Bio [\onlinecite{Schuemann2019:RR}].
For example, the reaction rate constant of $OH+H_2O_2 \rightarrow HO_2 + H_2O$ is given by
$k_{12}=0.0023\times 10^{10} /M/s = 0.023 /M/ns$.
Similarly the reaction rate constant of $OH + OH \rightarrow H_2O_2$, described by Eq. (\ref{Aeq1b}),
$d[H_2O_2]/dt = k_3 [OH]^2$, is $k_3=0.475\times 10^{10} /M/s = 4.75 /M/ns$.
In the absence of non-linearities ($k_3 = k_{12} = 0$) and zero diffusion, the linear rate constants, $k_1$ and $k_2$, can be determined from a steady-state condition where $u$ and $v$ are both constant so $du/dt = dv/dt = 0$, thus $v = (k_1/k_2) u$ where $G=0$.

With regards to differences with our latest work, presented in Ref. [\onlinecite{Baikalov2022}], we have omitted the labels for the track indices in $u$ and $v$ as the explicit inclusion of indices is convenient for the description of weak inter-track limit where the analytical solutions and the overlap integrals can be calculated perturbatively.
Nevertheless, to recover the rate equations in Ref. [\onlinecite{Baikalov2022}], we apply the following transformation $u = \sum_{i=1}^{N_s} u_i$ in Eq. (\ref{Aeq1a})
where $N_s$ denotes the number of particle tracks, identical to number of particles in a beam.
Substituting this transformation results in partitioning Eq. (\ref{Aeq1a}) into $N_s$ independent rate equations.
A one-to-one correspondence between the variables in this model and in Ref. [\onlinecite{Baikalov2022}] is the following: $u_i \rightarrow c_i$, $D_f \rightarrow \alpha$, $k_1 \rightarrow k_s$ and $2 k_3 \rightarrow k_r$.
The rest of parameters and variables, $k_2, k_{12}$ and $v$ were omitted in Ref. [\onlinecite{Baikalov2022}].

Therefore, the rate equations presented in the current study, Eqs. (\ref{Aeq1a}) and (\ref{Aeq1b}), are more general than their counterparts in Ref. [\onlinecite{Baikalov2022}]
and the solutions at the limit of weakly and/or strongly correlated tracks can be calculated non-perturbatively by numerical approaches such as finite difference and/or elements.

An interesting special limiting case of negligible $k_1$, $k_2$, $k_3$, and $k_{12}$ corresponds to an asymptotic solution of the Gaussian distribution function for $u$ at $t >> 0$ as studied in Ref. [\onlinecite{Baikalov2022}].
Note that we use slightly different initial condition such that at $t=0$ a constant distribution of ROS inside a cylinder with radius $w$ is considered, where $u = u_0$ within $r \leq w$ and zero otherwise.
And $v = v_0 = 0$ everywhere.
$w$ is the width of a particle track at initial time $t=0$ and it can be extracted from MC simulations of track structures of particles.
$w$ is a parameter that depends on particle LET.
The advantage of using this initial condition would be omission of parameter $\tau_0$ introduced in Ref. [\onlinecite{Baikalov2022}] in favor of the initial track width $w$.
Geometrically, tracks with this boundary condition do not suffer from spurious Gaussian tail at initial time.


At weak interaction limit we disregard the non-linear terms to calculate analytical solutions.
We further treat the non-linear terms perturbatively and calculate the corrections to linear solutions.
Note that the analytical solutions at the strong limit of non-linearities have been calculated in Ref.[\onlinecite{Abolfath2020:MP}].
Thus we may perform an interpolation between weak and strong interaction limits to calculate the solutions at the intermediate interacting limit where both linear and non-linear terms are comparable.
The rest of this presentation is devoted to calculate the solutions of these equations.

To handle the time dependence in the partial differential equations we perform a Laplace transformation
\begin{eqnarray}
\overline{u}(s, \vec{r}) = \int_{0}^{\infty} dt u(\vec{r}, t) e^{-s t},
\label{Aeq2}
\end{eqnarray}
with the inverse Laplace transformation, given by
\begin{eqnarray}
u(\vec{r}, t) = \frac{1}{2\pi i} \int_{\gamma-i\infty}^{\gamma+i\infty} dp \overline{u}(s, \vec{r}) e^{s t}.
\label{Aeq3}
\end{eqnarray}
Insertion of Eq. (\ref{Aeq2}) to the time-derivative term in Eqs. (\ref{Aeq1a}) and (\ref{Aeq1b}) yields
\begin{eqnarray}
\int_{0}^{\infty} dt \frac{\partial u(\vec{r}, t)}{\partial t} e^{-s t} = -u(0, \vec{r}) + s \overline{u}(s, \vec{r}),
\label{Aeq4}
\end{eqnarray}
where $u(0, \vec{r})$ can be specified by the initial condition for $u$ at $t=0$, $u(0, \vec{r}) = u_0(\vec{r})$.
Linearizing the rate equations and applying the Laplace transformation,
we treat the non-linear terms perturbatively
\begin{eqnarray}
s \overline{u}(s, \vec{r}) &=& \overline{G}(s) + D_f \nabla^2 \overline{u}(s, \vec{r}) \nonumber \\
&-& k_1 \overline{u}(s, \vec{r}) + k_2 \overline{v}(s, \vec{r}) + u(0, \vec{r})
\label{Aeq5}
\end{eqnarray}
and
\begin{eqnarray}
\overline{v}(s, \vec{r}) = \frac{k_1 \overline{u}(s, \vec{r}) + \overline{v}(0, \vec{r})}{s + k_2}
\label{Aeq6}
\end{eqnarray}
We can now replace Eq.(\ref{Aeq6}) in Eq.(\ref{Aeq5}) and reduce the system of coupled differential equations into a single equation in terms of $u$, thus
\begin{eqnarray}
\nabla^2 \overline{u}(s, \vec{r}) &-& q^2 \overline{u}(s, \vec{r}) = \nonumber \\
&-& \frac{1}{D_f}
\left[\frac{k_2 \overline{v}(0, \vec{r})}{s + k_2} + \left(\overline{G} + \overline{u}(0, \vec{r})\right)\right]
\label{Aeq7}
\end{eqnarray}
where
\begin{eqnarray}
q^2(s) = \frac{s}{D_f} \left(1 + \frac{k_1}{s + k_2}\right).
\label{Aeq8}
\end{eqnarray}
Applying the initial conditions everywhere
\begin{eqnarray}
u(0, \vec{r}) = v(0, \vec{r}) = 0
\label{Aeq9}
\end{eqnarray}
we find
\begin{eqnarray}
\nabla^2 \overline{u}(s, \vec{r}) - q^2 \overline{u}(s, \vec{r}) = - \overline{G}
\label{Aeq10}
\end{eqnarray}

Alternatively, we can start the time-evolution of the track expansion by applying the initial conditions right after entrance of the single track where we can consider $G=0$, from that time on.
Here the track structure insertion to the differential equations can be performed
through the boundary conditions $u(t=0, \vec{r}) = u_0 \theta(w - r)$, and $v(t=0, \vec{r}) = v_0 \theta(w - r)$, hence Eq. (\ref{Aeq7}) simplifies to
\begin{eqnarray}
\nabla^2 \overline{u}(s, \vec{r}) - q^2 \overline{u}(s, \vec{r}) = - V(s) \theta(w - r),
\label{Aeq11}
\end{eqnarray}
where
\begin{eqnarray}
V(s) = \frac{u_0}{D_f}\left[1 + \frac{k_2}{s + k_2}\frac{v_0}{u_0}\right].
\label{Aeq12}
\end{eqnarray}
Here $\theta(w - r)$ is a Heavyside function such that $\theta = 1$ if $r \leq w$ and zero, otherwise.
Eq. (\ref{Aeq11}) is of the general form given in Carslaw and Jaeger [\onlinecite{Carslaw1959:book}] for heat conduction between composite cylinders.
For $r \leq w$ the solutions of Eq. (\ref{Aeq11}) are
\begin{eqnarray}
\overline{u}_<(s, \vec{r}) = \frac{V}{q^2} - \alpha_1 I_0(qr),
\label{Aeq13}
\end{eqnarray}
\begin{eqnarray}
\overline{v}_<(s, \vec{r}) = \frac{k_1}{s + k_2} \overline{u}_<(s, \vec{r}) + \frac{v_0}{s + k_2}.
\label{Aeq13a}
\end{eqnarray}
And for $r > w$
\begin{eqnarray}
\overline{u}_>(s, \vec{r}) = \alpha_2 K_0(qr)
\label{Aeq14}
\end{eqnarray}
\begin{eqnarray}
\overline{v}_>(s, \vec{r}) = \frac{k_1}{s + k_2} \overline{u}_>(s, \vec{r})
\label{Aeq15}
\end{eqnarray}
where $\alpha_1(s)$ and $\alpha_2(s)$ are boundary matching parameters.
They are functions of Laplace transform variable $s$ and are determined by matching the boundary conditions across $r = w$.
At $r = w$, the continuity of the diffusion equation and their first derivatives imply
\begin{eqnarray}
\overline{u}_<(s, w)  = \overline{u}_>(s, w),
\label{Aeq16}
\end{eqnarray}
and
\begin{eqnarray}
\overline{u}'_<(s, w)  = \overline{u}'_>(s, w),
\label{Aeq17}
\end{eqnarray}
where $u'(s, w) = du(s, r)/dr$ at $r=w$.
Insertion of the boundary condition, Eqs. (\ref{Aeq16}) and (\ref{Aeq17}),
in Eqs. (\ref{Aeq13}) and (\ref{Aeq14}) solves for $\alpha_1$ and $\alpha_2$
\begin{eqnarray}
\alpha_1(s) = \frac{V}{q^2} \frac{1}{I_0(qw) + K_0(qw) I_1(qw) / K_1(qw)},
\label{Aeq18}
\end{eqnarray}
and
\begin{eqnarray}
\alpha_2(s) = \alpha_1 \frac{I_1(qw)}{K_1(qw)}.
\label{Aeq18}
\end{eqnarray}
Note that $\alpha_1$ and $\alpha_2$ are explicit function of $s$.
This is important in calculating the inverse transform of $u$ and $v$.

The interaction / non-linear terms must be treated perturbatively because upon Laplace transformation they turn to a
non-local integral in $s$.
For example applying Laplace transform over $u^2$ turns into an integral equation with two interacting fields
through a propagator
\begin{eqnarray}
\int_{0}^{\infty} dt u^2(\vec{r}, t) e^{-s t} &=& \frac{1}{(2\pi i)^2}
\int_{-i\infty}^{i\infty} ds' \overline{u}(\vec{r}, s') \nonumber \\ &&
\int_{-i\infty}^{i\infty} ds'' \overline{u}(\vec{r}, s'')
\frac{\theta(s - s' - s'')}{s - s' - s''}. \nonumber \\
\label{Aeq19}
\end{eqnarray}

In a weak non-linear coupling limit, we employ a perturbative approach to the non-linear terms in Eqs. (\ref{Aeq1a} and \ref{Aeq1b})
\begin{eqnarray}
\tilde{u} = u + u',
\label{Aeq20a}
\end{eqnarray}
and
\begin{eqnarray}
\tilde{v} = v,
\label{Aeq20b}
\end{eqnarray}
where $u$ and $v$ are the solutions of the linear equations of Eqs. (\ref{Aeq1a} and \ref{Aeq1b}) where $k_3 = k_{12} = 0$.
An equation for $u_2$ can be derived after substituting Eqs. (\ref{Aeq20a} and \ref{Aeq20b}) into Eqs. (\ref{Aeq1a} and \ref{Aeq1b})
\begin{eqnarray}
u' = - \frac{k_3 u^2}{k_1 + 2k_3 u}
\label{Aeq21}
\end{eqnarray}

\subsection{Useful identities}
Useful identities (see page 375, Eq. (9.6.15), Ref. [\onlinecite{Abramowitz1964:book}])
\begin{eqnarray}
I_\nu(z) K_{\nu+1}(z) + K_\nu(z) I_{\nu+1}(z) = \frac{1}{z},
\label{Aeq22}
\end{eqnarray}
thus we find
\begin{eqnarray}
I_0(qw) K_1(qw) + K_0(qw) I_1(qw) = \frac{1}{qw}
\label{Aeq22}
\end{eqnarray}

\subsection{Fourier transform}
Fourier transform
\begin{eqnarray}
f(\omega) = \frac{1}{\sqrt{2\pi}} \int_{-\infty}^{\infty} dt f(t) e^{i\omega t}
\label{Aeq23}
\end{eqnarray}
Inverse Fourier transform
\begin{eqnarray}
f(t) = \frac{1}{\sqrt{2\pi}} \int_{-\infty}^{\infty} d\omega f(\omega) e^{-i\omega t}
\label{Aeq24}
\end{eqnarray}
The following identity will be used for the inverse Laplace transform expression
\begin{eqnarray}
f(t') = \frac{1}{2\pi}
\int_{-\infty}^{\infty} d\omega e^{-i\omega t'}
\int_{-\infty}^{\infty} dt e^{i\omega t} f(t)
\label{Aeq25}
\end{eqnarray}
The delta-function
\begin{eqnarray}
\delta(t-t') = \frac{1}{\sqrt{2\pi}} \int_{-\infty}^{\infty} d\omega e^{-i\omega (t-t')}
\label{Aeq26}
\end{eqnarray}
\begin{eqnarray}
\delta(\omega - \omega') = \frac{1}{\sqrt{2\pi}} \int_{-\infty}^{\infty} dt e^{i(\omega-\omega') t}
\label{Aeq27}
\end{eqnarray}

\subsection{Laplace transform}
Laplace transform
\begin{eqnarray}
f(s) = {\cal L}[F(t)](s) = \int_{0}^{\infty} dt F(t) e^{-s t}
\label{Aeq28}
\end{eqnarray}
Inverse Laplace Transform – Bromwich Integral
\begin{eqnarray}
F(t) = {\cal L}^{-1}[f(s)](t)
\label{Aeq29}
\end{eqnarray}
We define (see page 908 in Ref. \onlinecite{Arfken1995:book})
\begin{eqnarray}
F(t) = e^{\gamma t} G(t)
\label{Aeq30}
\end{eqnarray}
Note $t \geq 0$.
If $F(t)$ diverges as $e^{\alpha t}$, it is required $\gamma$ to be greater than $\alpha$.
From Fourier transform, Eq.(\ref{Aeq25}) we have ($t \rightarrow t'$ and $t' \rightarrow t$ and considering only a domain of positive time, $t \geq 0$):
\begin{eqnarray}
G(t) = \frac{1}{2\pi}
\int_{-\infty}^{\infty} d\omega e^{i\omega t}
\int_{0}^{\infty} dt' e^{-i\omega t'} G(t').
\label{Aeq31}
\end{eqnarray}
Introducing a complex variable $s = \gamma + i\omega$, and replacing $i\omega = s - \gamma$, and $ds = id\omega$ (assuming $\gamma$ is a constant), in Eq.(\ref{Aeq31}) we find
\begin{eqnarray}
G(t) = \frac{1}{2\pi i}
\int_{\gamma-i\infty}^{\gamma+i\infty} ds e^{(s - \gamma) t}
\int_{0}^{\infty} dt' e^{-(s - \gamma) t'} G(t').
\label{Aeq32}
\end{eqnarray}
Introducing $F(t) = e^{\gamma t} G(t)$
\begin{eqnarray}
G(t) &=& \frac{1}{2\pi i}
\int_{\gamma-i\infty}^{\gamma+i\infty} ds e^{(s - \gamma) t}
\int_{0}^{\infty} dt' e^{-s t'} F(t') \nonumber \\ &=&
\frac{1}{2\pi i}
\int_{\gamma-i\infty}^{\gamma+i\infty} ds e^{(s - \gamma) t} f(s).
\label{Aeq33}
\end{eqnarray}
Finally the integral expression for the inverse Laplace transform is given by
\begin{eqnarray}
F(t) = G(t) e^{\gamma t} &=&
\frac{1}{2\pi i}
\int_{\gamma-i\infty}^{\gamma+i\infty} ds e^{s t} f(s) \nonumber \\
&=& {\cal L}^{-1}[f(s)](t).
\label{Aeq33}
\end{eqnarray}


\begin{figure}
\begin{center}
\includegraphics[width=1.0\linewidth]{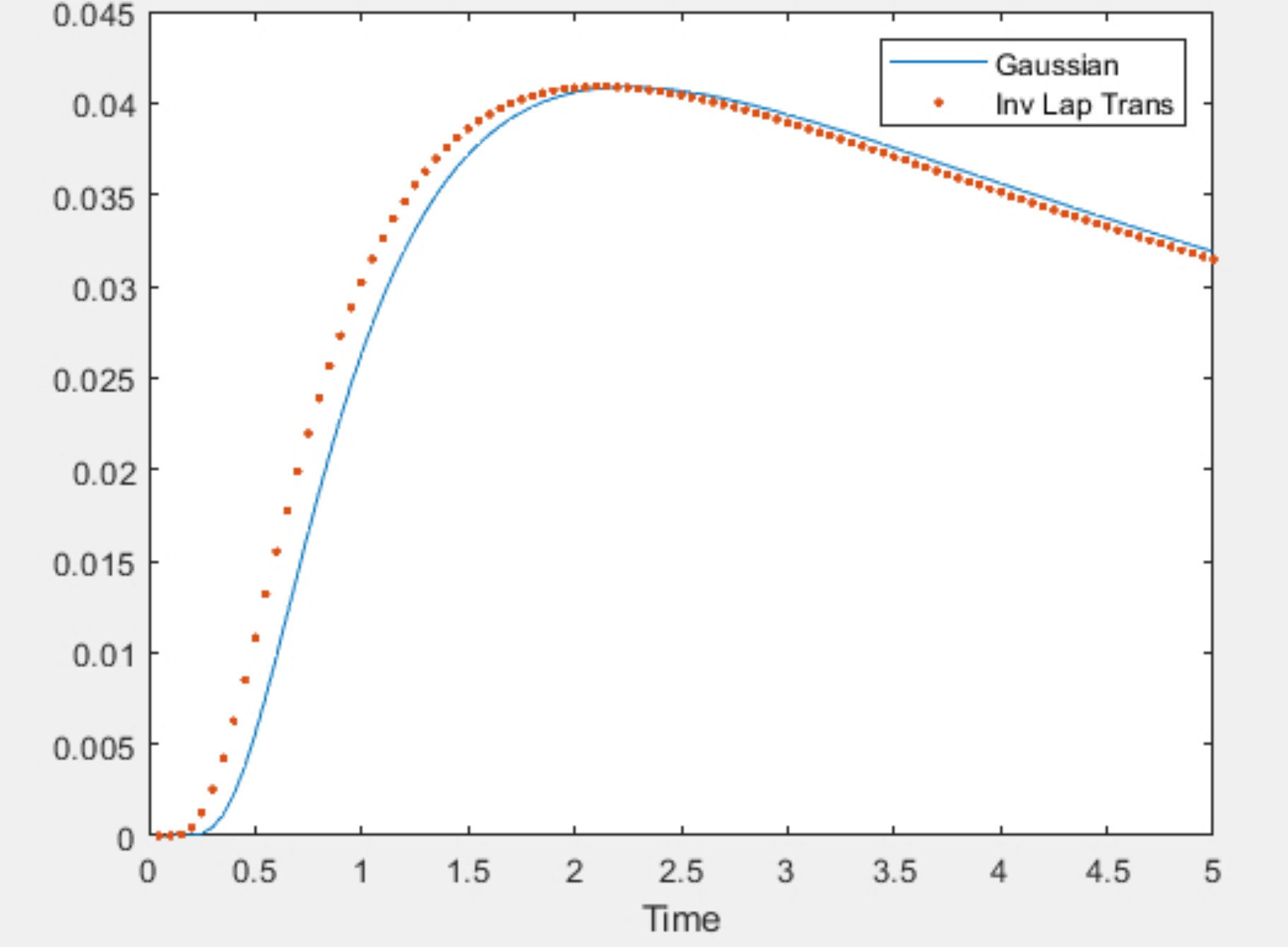}\\ 
\noindent
\caption{
Numerical solution of $u_>(\vec{r}, t)$ calculated by inverse Laplace transform (red dots) and a Gaussian function (solid line), given by Eq. (\ref{Aeq34}), are shown. At large distances, i.e., $r >> w$, Gaussian is an approximate fit to $u_>(\vec{r}, t)$.
The numerical values of the parameters used in the integration to produce this figure are $w=1, r=3, D_f=1, k_1=k_2=k_3=k_{12}=0$.
}
\label{fig5}
\end{center}\vspace{-0.5cm}
\end{figure}

\subsection{Appendix: Gaussian solutions}
\label{AppV}
In a limit of ideal diffusion (in the absence of all reaction rates) the asymptotic solutions of $u_>(r, t)$ for large arguments, $r >> w$, follow Gaussian distributions
multiplied by the initial number of chemical species, $\pi w^2 u_0$
\begin{eqnarray}
u_>(\vec{r}, t) = \frac{\pi w^2 u_0}{4\pi D_f t} e^{-\frac{r^2}{4D_f t}}.
\label{Aeq34}
\end{eqnarray}
This can be obtained from calculation of invrese Laplace transform of $u_> (s, r) = \alpha_2(s) K_0(q(s) r)$
\begin{eqnarray}
{\cal L}^{-1}[u_>(qr)](t) &=& \frac{1}{2\pi i} \int_{\gamma-i\infty}^{\gamma+i\infty} ds e^{s t} u_> (s, r)  \nonumber \\
&=& \frac{1}{2\pi i}  \int_{\gamma-i\infty}^{\gamma+i\infty} ds e^{s t} \alpha_2(s) K_0\left(q(s) r\right). \nonumber \\
\label{Aeq35}
\end{eqnarray}
Calculation of this integral requires numerical integration of Eq.(\ref{Aeq35}),
recalling the explicit dependence of $\alpha_2$ and $q$ on Laplace transform variable, $s$.
We have performed this calulation and verified validity of Eq.(\ref{Aeq34}) as illustrated in Fig. (\ref{fig5}).
In this figure, the numerical integration of inverse Laplace transform of $u_>(\vec{r}, t)$, and fitting to a Gaussian PDF as given in Eq. (\ref{Aeq34}) are plotted.


\end{document}